\documentclass[prb,preprint]{revtex4-2} 
\usepackage[utf8]{inputenc}  
\usepackage[T1]{fontenc}  
\usepackage{helvet}
\usepackage{amsmath}
\usepackage{graphicx}
\usepackage{amssymb} 
\usepackage{url}
\usepackage{braket}
\usepackage{xspace}

\newcommand{\SO}{\mathrm{SO}}
\newcommand{\SU}{\mathrm{SU}}
\newcommand{\hspin}{spin-$1/2$\xspace}

\begin{document}
\title{The spinorial ball: a macroscopic object of \hspin}
\author{Samuel Bernard-Bernardet}
\affiliation{DotWave Lab, Chamb\'{e}ry, France}
\author{David Dumas}
\affiliation{University of Illinois at Chicago, USA}
\author{Benjamin Apffel}
\email{benjamin.apffel@epfl.ch}
\affiliation{Laboratory of Wave Engineering, EPFL, Lausanne, Switzerland}
\date{June 2023}
\begin{abstract}
The goal of this article is to introduce a new object named the spinorial ball, with the goal of providing insight on how half-integer spin behave. The later is generally introduced to students in quantum physics lectures as ``an object that does not come back to its original state after one full turn but that does after two''. Due to this context, it is often thought and considered as a purely quantum behavior. However, \hspin is more than anything a geometrical property of the rotations group, that can also have concrete consequences at a macroscopic scale. Therefore, we propose here to describe and introduce \hspin without the help of quantum mechanics but rather through geometry. We also introduce several general group theory concepts such as group homomorphisms and homotopy classes of paths through the example of the groups $\SU(2)$ and $\SO(3)$. The spinorial ball provides a macroscopic visualization of all these concepts, which are ubiquitous in quantum physics.
\end{abstract}
\maketitle
\section{Introduction}
\label{sec:intro}
\subsection{Foreword and context}
  We propose here to describe and introduce \hspin without the help of quantum mechanics but rather through geometry and a bit of group theory. We believe that this approach will help students and interested readers to get better insight on it. As it has been done in many good textbooks, we will not discuss here the particular (and important) manifestation of \hspin in quantum mechanics. The interested reader can for instance refer to \citep{basdevant_quantum_2005} for a compact introduction and to \citep{cohen-tannoudji_quantum_1986} for a more technical approach. Although the authors have tried to make this presentation self-contained, basic knowledge about rotations and matrix theory could help for some technical aspects (see for instance \cite{appel_mathematics_2007}). Whenever possible we focus on the underlying concepts and physical intuitions, at times omitting technical details for the sake of such focus.  We provide references for proofs that are omitted for this reason. Last, we would like to point out that the name of spinorial ball initially originates from an article from J.~Baez and J.~Huerta dedicated to visualisation of other Lie groups \citep{baez2014g2}.
 
\subsection{Macroscopic \hspin}
\begin{figure}
\centering
\includegraphics[width=16cm]{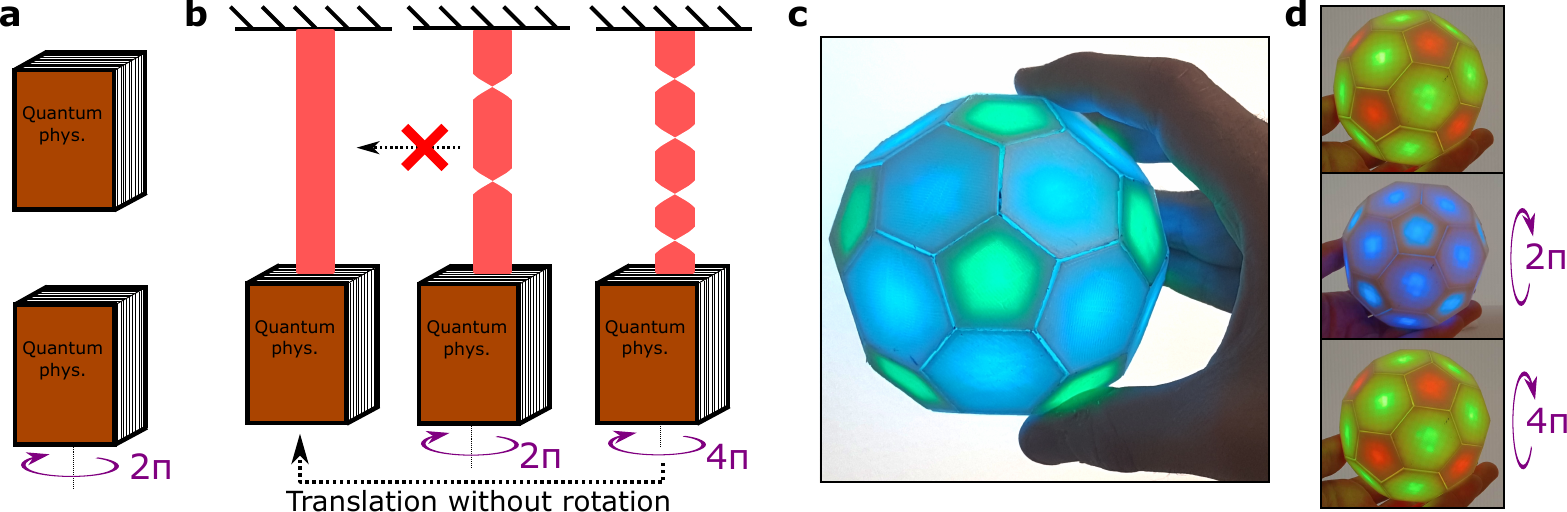}
\caption{(a) Any unconstrained physical object comes back to its initial state after a rotation of $2\pi$ around any axis. (b) When a ribbon fixed to a wall is added, one can not remove the twist in this ribbon after a rotation of $2\pi$ without further rotating the object. After a second turn, the twists can be removed without any additional rotation. (b) Picture (with removed background) of the spinorial ball. When the ball rotates, the colors displayed on the faces change continuously. (d) Starting from any initial state (top), a rotation of $2\pi$ around any axis brings the ball to another state (middle) but a rotation of $4\pi$ brings it back to its initial state (bottom).}
\label{fig:1b}
\end{figure}

If one takes an object, say a favorite quantum mechanics book, rotating it by a full turn around any axis will return it to its original state as in Fig. \ref{fig:1b}a). On the contrary, \hspin objects are entities that are not invariant by this operation, as two complete turns are needed to bring them back to their initial state. This unusual features makes them puzzling for the intuition. As many teachers know, it is possible to create analogs of \hspin at a macroscopic scale by adding some constraints on a physical object. For instance, one can tie the book to one end of a ribbon, the other end being clamped to a wall as in Fig. \ref{fig:1b}b. After one complete rotation of the book, the ribbon acquires a twist that cannot be removed without rotating the object back. If one applies a second turn to the book, the situation seems even worse as more twists appear on the ribbon. However, those twists can be removed `magically' without rotating the book by using a combination of translations (see for instance videos   
\cite{HiseBeltTrickVideo}
\cite{alex_mason_httpswwwyoutubecomwatchvnat-esrextq_2008}). 
This is nothing but the famous \textit{Dirac belt trick} (cf.~\cite[pp.~43-44]{penrose_rindler_1984}) that is widely used in quantum mechanics lectures to help students understand how \hspin behaves. Indeed, it  exhibits behavior in which $2\pi$ rotation does not leave the system \{book + ribbon\} invariant while a $4\pi$ does, at least up to translations. 

There are nevertheless several issues with this picture. The translations needed to restore the original state of the system have no direct analog in quantum mechanics (and even seem like a bit of `cheating'). One can also wonder if there is anything analogous to the constraints introduced in this trick for free-to-move quantum particles. Although useful, this picture is therefore somewhat limited as a way to provide better understanding of \hspin.

Another macroscopic \hspin system based on coupled mechanical pendulums has also been described \cite{leroy_simulating_2006, leroy_simulating_2010}, but the effective \hspin that emerges takes place in an abstract space of parameters. Therefore, a direct and unconstrained physical realization of a half-integer spin at macroscopic scale is still missing. We now propose to fill this gap by introducing the \emph{spinorial ball}.

\subsection{First contact with the ball}
 The spinorial ball is a polyhedron of roughly spherical shape with pentagonal and hexagonal faces (see picture in Fig. \ref{fig:1b}). Each face is illuminated with LEDs in a certain color (all the possible colors are shown in Fig. \ref{fig:3}a). In this section we explain the main observations likely to result from casual experimentation with the ball by non-experts.   While we primarily focus on its implementation as a manipulable electronic device, a mobile version \cite{spinPhone} (which works by moving the phone) and a browser-based simulation \cite{noauthor_httpproxy-informatiquefrquball_nodate} are available, and we encourage the reader to use them as an experimental apparatus while reading this introduction.

If one starts to move the ball, the colors of the faces change smoothly. Quickly, one notices two things:
\begin{itemize}
    \item All pentagons (resp.~all hexagons) are illuminated in the same color, which we denote $C_P$ (resp.~$C_H$). 
    \item The colors displayed do not change when the ball is translated, but they do when the ball is rotated.
\end{itemize}
The ball being only sensitive to rotations, it is natural to look more closely at the way the colors behave with respect to them. The ball being a very symmetrical object, its orientation is actually somewhat difficult to track visually, but it can be done.  Some kind of temporary asymmetric decoration could also be added, such as a sticker or piece of tape marking a reference point.

After moving the ball in various ways, it becomes evident that no two orientations of the ball correspond to the same pair of colors $(C_P,C_H)$. Therefore, the colors displayed on the faces seem to provide complete information about the rotation applied. To make sure that this is correct, one might start from an initial state as in Fig. \ref{fig:1b}d and perform a full turn of the ball about an axis.   In that process, the ball does not display the same combination of colors twice, as expected from the previous remark. However, once the ball is back to its original orientation after a full turn, the pentagons and the hexagons do not display their initial colors (see Fig. \ref{fig:1b}d), meaning that the system \{ball + colors\} is not back to its initial state, although the system \{ball\} alone is.  It is only after a second full turn that the colors return to their original state. The colors therefore seem to have replaced the ribbon from the Dirac belt trick. 

After further experiments, one may realize that any given orientation of the ball can only every correspond to two specific color states, regardless of rotations that may occur in the meantime. After a bit of work, optimistically one might gather enough evidence to (correctly) hypothesize:
\begin{itemize}
    \item A specific set of colors corresponds to only one orientation of the ball. Therefore, the observation of the colors is enough to determine the current orientation of the ball.
    \item The converse of the previous property fails:  For each physical orientation of the ball, there are two sets of colors $(C_{P1}, C_{H1})$ and $(C_{P2}, C_{H2})$ that can be displayed.  Which one appears depends on the history of rotations applied to the ball thus far.
    \item For a given orientation, one can go from one set of colors to another by performing a full turn rotation around any axis. Continuing to apply full rotations about an axis, the set of colors displayed alternates between one and the other, thus showing whether an odd or an even number of turns have occurred.
\end{itemize}

The fact that the orientation of the ball is encoded in the color state means that any additional asymmetric decoration is actually unnecessary, and that the symmetry of the object is a natural design choice.  Observation of the colors $(C_P, C_H)$ alone can be used to determine the ball's orientation, and asymmetry of the object would only provide a redundant source of orientation information. Although we know that the `physical' ball (ignoring illumination) is back in its initial state after one turn, the symmetric physical structure and different color palette mean this is not easily noticed. One can only observe that the ball comes back to its initial state after two turns. The illuminated ball therefore acts as a free-to-move macroscopic \hspin object. From this, several questions can be raised. What do the colors represent? How exactly are the colors related to the rotations of the ball? Why do the colors behave like a \hspin? Could we create a spin-1/3 in a similar way, encoding each orientation in three possible color schemes?

In the subsequent sections, we answer these question by carefully explaining how the spinorial ball works.  We first introduce the rotation group $\SO(3)$ to describe the possible rotations of the physical ball. We then introduce the spinors and the group $\SU(2)$, which will correspond to the the different illumination states of the ball.  We then define a group homomorphism between $\SU(2)$ and $\SO(3)$ that will provide the link between the physical rotation of the ball (an element of $\SO(3)$) with the colors it displays (an element of $\SU(2)$). Having introduced all the needed concepts, we come back in the last section to the spinorial ball itself and discuss the question raised above.

\section{Rotation group}
\subsection{Definition}
The possible rigid motions of an object that preserve handedness and do not move a chosen center point are elements of the \textit{rotation group} in $\mathbb{R}^3$ that will be denoted $\SO(3)$. Here, the center point is taken as the center of the spinorial ball and the elements of $\SO(3)$ will correspond to all the possible rotations of the ball that we will now describe mathematically. Formally, the rotation group is defined as the set of $3 \times 3$ real matrices $R$ (or linear transformations of $\mathbb{R}^3$, thus preserving the origin) that
\begin{itemize}
\item[(P1)] preserve the length of all vectors : $R^\dag R = \mathbb{I}_3$
\item[(P2)] preserve the orientation of space : $\det R = 1$
\end{itemize}

One can verify that $\SO(3)$ is a non-commutative group under the operation of matrix product (or composition of motions), meaning that the product and the inverse of a rotation is also a rotation. Moreover, one can show that any element $R \neq \mathbb{I}_3$ of $\SO(3)$ admits a unique (up to sign) unit vector $\vec n$, called its \textit{rotation axis}, such that $\vec n$ is invariant under the rotation: $R \vec n = \vec n$. In the plane orthogonal to $\vec n$, one can moreover show that $R$ acts as two-dimensional rotation through some angle $\theta$.  This means that all the elements of $\SO(3)$ correspond to rotations of space, justifying the name ``rotation group'' for $\SO(3)$. Conversely, any physical rotation of space with an angle $\Psi$ around the axis $\vec n$ can be encoded in an element of $\SO(3)$ that we write $R_{\vec n} (\Psi)$. For instance, the rotation of angle $\Psi$ around the $x$-axis (that is, taking $\vec n = \vec x$) is
\begin{equation}
R_{\vec x} (\Psi) = \begin{bmatrix}
1 && 0 && 0 \\
0 && \cos{\Psi} && -\sin{\Psi} \\
0 && \sin{\Psi} && \cos{\Psi} \\
\end{bmatrix}
\label{eq:rotz}
\end{equation}
which satisfies (P1) and (P2) and therefore belongs to $\SO(3)$.

\subsection{Generators of rotations}
There exists a convenient way to specify any rotation, i.e.~any element of $\SO(3)$. Let us consider first an infinitesimal rotation $d\Psi$ around one of the coordinate unit vectors $\vec x$, $\vec y$, $\vec z$. 
These rotation matrices can be written in the form $R_{\vec n}(d \Psi) = \mathbb{I}_3 - i\, d\Psi J_n$ with 
\begin{equation}
J_x = \begin{bmatrix}
0 && 0 && 0 \\
0 && 0 && -i \\
0 && i && 0 \\
\end{bmatrix}
\quad
J_y = \begin{bmatrix}
0 && 0 && i \\
0 && 0 && 0 \\
-i && 0 && 0 \\
\end{bmatrix}
\quad
J_z = \begin{bmatrix}
0 && -i && 0 \\
i && 0 && 0 \\
0 && 0 && 0  \quad
\end{bmatrix}
\end{equation}
More precisely, these expressions give the derivatives of $R_{\vec n}(\Psi)$ as a function of $\Psi$, evaluated at $\Psi=0$.  (The expression for $J_x$ follows directly from \eqref{eq:rotz} above.)

To generate a macroscopic rotation of angle $\Psi$ around $\vec x$, one can compose $N \gg 1$ rotations of $\Psi/N$, and for $N$ large each of these factors should be close to the infinitesimal rotation considered above. 
 This leads to the expression $R_{\vec{x}}(\Psi) = \lim_{N \to\infty} \left( \mathbb{I}_3 - i\frac{\Psi}{N} J_x \right)^N = e^{-i \Psi J_x}$, the exponential being taken in the sense of matrices. In the same manner, two matrices $J_y$ and $J_z$ that can be used to generates the rotations around $\vec y$ and $\vec z$ respectively. More generally, any rotation of angle $\Psi$ around a unitary axis $\vec n$ can be written as 
\begin{equation}
R_{\vec n} (\Psi) = e^{-i\Psi \vec J \cdot \vec n}
\label{eq:expRot}
\end{equation}
where $\vec J = (J_x, J_y, J_z)$.
Linear combinations of the three matrices $ -i \vec J = (-iJ_x, -iJ_y, -iJ_z)$ are thus enough to generate all the rotations through the exponential, and so these matrices are called \textit{generators} of $\SO(3)$. In more mathematical terminology, the three matrices of $-i \vec J$ would be called a basis of the Lie algebra $\mathfrak{so}(3)$. 

The generators satisfy the commutation relations
\begin{equation}
[J_x, J_y] = i J_z, \quad [J_y, J_z] = i J_x, \quad [J_z, J_x] = i J_y 
\label{eq:comRot}
\end{equation}
where $[A,B] = AB - BA$ refers to the matrix commutator, measuring failure of commutativity for a pair of matrices. Those relations are also encountered in quantum mechanics for angular momentum description. This means that up to a $\hbar$ factor, the generators of rotations can be identified with angular momentum operators considered in quantum mechanics. This is the angular counterpart of the fact that that linear momentum arises from the generators of the group of translations.

The rotation group provides a clear mathematical description of the ball's physical motion. We will now formally introduce the `color space' of the ball, and see the analogue of rotations for it.

\section{Rotation and spin}
\subsection{Spin-1/2 and $\SU(2)$}
\label{subsec:spinSU2}
A pair of colors $(C_P, C_H)$  displayed on the ball encodes a unit vector $s$ of $\mathbb{C}^2$, with $C_P$ determined by one component and $C_H$ by the other through a color mapping function.
A variety of mappings from complex numbers to colors might be used, but it is essential that every unit vector corresponds to a different ordered pair of colors.
The color mapping chosen in our implementation is detailed in section \ref{subsec:generalPrinc}, and with such a mapping fixed, we can blur the distinction between components of a vector and their corresponding colors.  Following conventions common in introductory quantum mechanics, we thus denote the vector corresponding to a pair of colors by
\begin{equation}
s=C_P \ket{\uparrow} + C_H \ket{\downarrow}, \quad (C_P, C_H) \in \mathbb{C}^2, \quad |C_P|^2 + |C_H|^2 = 1
\end{equation}
where $\ket{\uparrow}$ and $\ket{\downarrow}$ are basis vectors that would correspond to up and down spin states in quantum mechanics. In the following, any such unit vector in $\mathbb{C}^2$ will be called a spinor, but should be thought as an illumination state of the ball. 

We now aim to define the equivalent of rotations in this color space. Inspired by the previous section, one can consider all the $2 \times 2$ complex matrices $S$ acting on $\mathbb{C}^2$ that satisfy the two-dimensional complex analogues of (P1) and (P2). This set of matrices is once again a group that is usually denoted $\SU(2)$, and is the analogue of rotations for spinors. After a bit of algebra, one can show that each element $S$ of $\SU(2)$ has the form
\begin{equation}
S = \begin{bmatrix}
a && -b^* \\
b && a^* \\
\end{bmatrix}, \quad (a, b) \in \mathbb{C}^2, \quad |a|^2 + |b|^2 = 1
\label{eq:SU2}
\end{equation} 
and conversely, any matrix of this form belongs to $\SU(2)$. Note that there exists a bijective correspondence between spinors and elements of $\SU(2)$ in which spinor $(C_P,C_H)$ corresponds to the matrix above with $(a,b) = (C_P,C_H)$.  As we aim to describe how colors `rotate' when the physical ball is rotated, we now focus on $\SU(2)$. 

\subsection{Generators of $\SU(2)$}
As for rotations in $\SO(3)$, any element of $\SU(2)$ can be written in exponential form as
\begin{equation}
S_{\vec n} (\Psi) = e^{-i \Psi \vec S \cdot \vec n}
\label{eq:expSpin}
\end{equation}
where $\Psi$ is a real number, $\vec n$ is a unit vector of $\mathbb{R}^3$, and where we have defined three generators of $\SU(2)$, denoted $\vec S = (S_x, S_y, S_z)$, by
\begin{equation}
 S_x = \frac{1}{2} \begin{bmatrix}
0 && 1 \\
1 && 0 \\
\end{bmatrix}
\quad S_y = \frac{1}{2} \begin{bmatrix}
0 && -i \\
i && 0 \\
\end{bmatrix}
\quad S_z = \frac{1}{2} \begin{bmatrix}
1 && 0 \\
0 && -1 \\
\end{bmatrix}
\label{eq:genSU2}
\end{equation}
Up to a factor $1/2$, these are also the \emph{Pauli matrices} of quantum mechanics. They obey the commutation relations
\begin{equation}
[S_x, S_y] = i S_z, \quad [S_y, S_z] = i S_x, \quad [S_z, S_x] = i S_y.
\label{eq:comSpin}
\end{equation}

The commutation relations (\ref{eq:comRot}) and (\ref{eq:comSpin}), combined with the formulas (\ref{eq:expRot}) and (\ref{eq:expSpin}) suggest a strong connection between $\SO(3)$ and $\SU(2)$. The next section describes the link between the two groups, which is at the heart of the spinorial ball.

\subsection{Group homomorphism between $\SU(2)$ and $\SO(3)$}

The previous considerations suggest a relation between $\SU(2)$ and $\SO(3)$ might involve identifying generators $S_x, S_y, S_z$ with $J_x, J_y, J_z$ respectively. The formulas (\ref{eq:expRot}) and (\ref{eq:expSpin}) suggest more generally that one can associate an element of $\SO(3)$ to an element of $\SU(2)$ through the map
\begin{equation}
\begin{array}{l|rcl}
T : & SU(2) & \longrightarrow & \SO(3) \\
    &S_{\vec n} (\Psi) =e^{-i \Psi \vec S \cdot \vec n} & \longmapsto & R_{\vec n} (\Psi) =e^{-i \Psi \vec J \cdot \vec n} \end{array}
\end{equation}

\begin{figure}
\centering
\includegraphics[width=16cm]{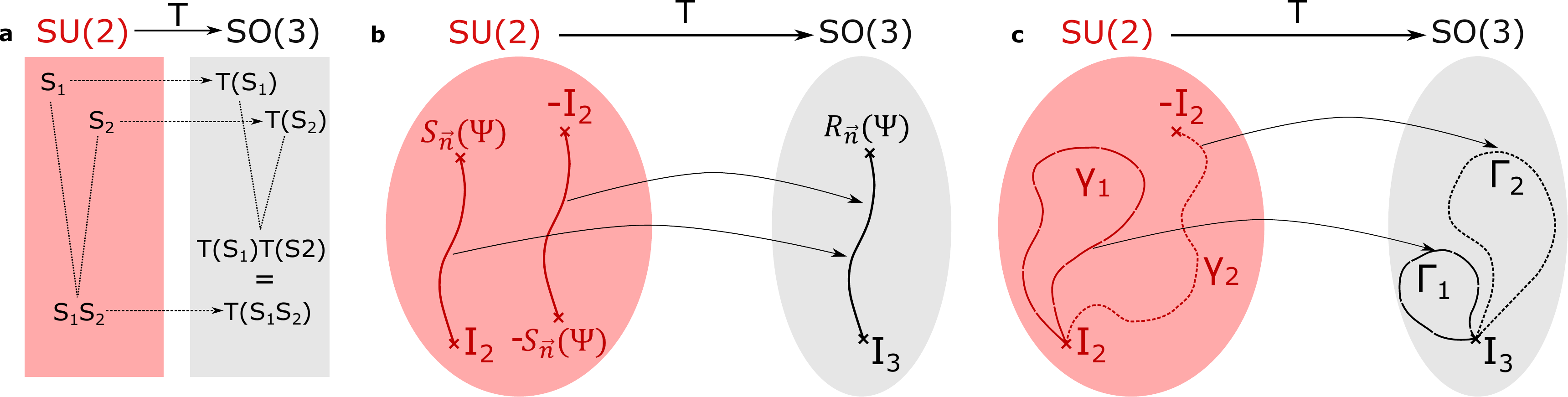}
\caption{(a) The map $T$ is a group homomorphism: The image of a product is the product of images, i.e. $T(S_1)T(S_2) = T(S_1 S_2)$.  (b) The group homomorphism $T$ maps $\SU(2)$ onto $\SO(3)$, but each elements of $\SO(3)$ is the image of two elements in $\SU(2)$.  In fact, any continuous path of $\SO(3)$ is the image of two continuous paths in $\SU(2)$, its \emph{lifts}. (c) A loop $\Gamma_1$ in $\SO(3)$ which is continuously contractible to a point has a lift with the same property, but a non-contractible path (e.g. $\Gamma_2$) may lift to an arc that joints two opposite points of $\SU(2)$.}
\label{fig:2}
\end{figure}

Indeed, this map is well-defined and has the important property that it is compatible with the group structure: for any $S_1, S_2$ in $\SU(2)$, one has $T(S_1 S_2) = T(S_1) T(S_2)$ in $\SO(3)$ \cite{appel_mathematics_2007}. In other words, taking the product of two elements in $\SU(2)$ and applying $T$ is the same as first applying $T$ to both of them and then taking the product in $\SO(3)$ (see Fig. \ref{fig:2}a). Due to this property, the map $T$ is said to be a \textit{group homomorphism} from $\SU(2)$ to $\SO(3)$.

However, $T$ can not be inverted as it is not injective: In $\SO(3)$, $R_{\vec n} (\Psi)$ is a rotation by angle $\Psi$ around $\vec n$. Therefore, if one chooses $\Psi  + 2 \pi$ instead of $\Psi$, the same rotation is obtained, i.e.
\begin{equation}
R_{\vec n}(\Psi+2\pi) = R_{\vec n} (\Psi) 
\end{equation}
However, this is not true of the parameterization of $\SU(2)$ introduced above. In fact, one can show that the Rodrigue-Euler formula holds
\begin{equation}
S_{\vec n} (\Psi) = \cos{\left( \frac{\Psi}{2} \right)} \mathbb{I}_2 - 2 i \sin{\left( \frac{\Psi}{2} \right) } \vec S \cdot \vec n 
\label{eq:RodrigueEuler}
\end{equation}
so that in particular
\begin{equation}
S_{\vec n} (\Psi + 2\pi) = -S_{\vec n} (\Psi).
\label{eq:minusSign}
\end{equation}
Thus $\pm S_{\vec n}(\Psi)$ are associated through $T$ to the same rotation
\begin{equation}
T(S_{\vec n} (\Psi + 2\pi)) = R_{\vec n}(\Psi+2\pi) = R_{\vec n} (\Psi) = T(S_{\vec n} (\Psi)).
\end{equation}
Stated another way, we see that taking $\Psi \rightarrow \Psi + 2\pi$ has no effect on the rotation in $\SO(3)$ but changes the sign in $\SU(2)$. This is of course directly linked to the \hspin properties discussed in section \ref{sec:intro}. One can moreover show that each element of $\SO(3)$ arises from \emph{exactly} two elements of $\SU(2)$ (differing by sign), and $T$ is therefore called a \textit{double cover}.

Obtaining a non-invertible map is more than just an artifact of our construction; it reflects a genuine difference between the groups $\SU(2)$ and $\SO(3)$.  These two spaces are topologically different, and there is no continuous bijection between them. This is discussed further in section \ref{subsec:pathSO3}.

\subsection{Path lift}
\label{subsec:pathLift}
As each element of $\SO(3)$ admits two preimages in $\SU(2)$, it is impossible to invert the map $T : \SU(2) \rightarrow \SO(3)$. Therefore, acting on the spinor (that is, the colors) by rotating the ball seems to be ill-defined. This is correct from a global point of view, but one can still perform an inversion process when the rotation is given not as a single object but as a smooth process. And that is physically natural, as in reality we do not apply a given rotation instantaneously.  Instead, the process of movement itself gives a continuous family of rotations, which we refer to as a \textit{path}. Formally, it is defined as a continuous function $\Gamma : t \in [0, 1] \mapsto R(t) \in \SO(3)$, i.e.~a continuous one-parameter family of rotation matrices.  The elements $\Gamma(0)$ and $\Gamma(1)$ are the endpoints of that path.

One can for instance consider $\Gamma(t) = R_{\vec n} (\Psi t)$ that goes continuously from $\Gamma(0) = \mathbb{I}_3$ to $\Gamma(1)=R_{\vec n} (\Psi)$, which corresponds physically to continuously rotating the ball about $\vec n$ until the total rotation angle reaches $\Psi$. As discussed before, each element of such a path has two preimages $\pm S_{\vec n}(\Psi t)$ in $\SU(2)$.  Moreover, there exist exactly two continuous paths in $\SU(2)$, namely $\{S_{\vec n} (\Psi t) \}_{0\leq t \leq 1}$ and $\{ -S_{\vec n} (\Psi t) \}_{0\leq t \leq 1}$, that map to $\Gamma$ under $T$.  This is shown in Fig. \ref{fig:2}b. Those paths go from $\mathbb{I}_2$ to $S_{\vec n}(\Psi)$ and from $-\mathbb{I}_2$ to $-S_{\vec n}(\Psi)$ respectively. If one specifies $\mathbb{I}_2$ as the starting point in $\SU(2)$, then there exists only one way to continuously lift a path from $\SO(3)$ to $\SU(2)$. In other words, there exists a unique continuous way to associate a preimage of \textit{each} element of this path once the preimage of \textit{one} element of the path is chosen. 

One can generalize this result to show that the same holds for any continuous path in $\SO(3)$. Namely, such a path in $\SO(3)$ has two associated paths (lifts) in $\SU(2)$, and choosing a preimage for one point along the path in $\SO(3)$ suffices to distinguish one of them. The path liftin property will be the key ingredient to realize a spinorial ball, as it provides a recipe to go from the physical rotation of the ball to an action on the colors.

\subsection{Homotopy classes of paths in $\SO(3)$}
\label{subsec:pathSO3}
Before coming back to the spinorial ball, we briefly discuss how the lifting of paths from $\SO(3)$ to $\SU(2)$ gives some insight into why these spaces are topologically different. For this we will consider \textit{loops}, i.e.~paths with the same starting and ending point: $\Gamma(0) = \Gamma(1)$. We then define the \textit{homotopy classes of paths} as collections of loops that can be continuously deformed to one to another. Note that in the rigorous definition of \textit{deformation}, the length of the path is irrelevant and the path itself should be thought of intuitively as an infinitely elastic rope that can be contracted or extended at will and which can furthermore pass through itself without getting snagged or tangled. For instance, in the plane $\mathbb{R}^2$ any loop can be continuously deformed into any another by suitable pushing and stretching, as the plane offers no obstacles to such movement.  There is therefore only one homotopy class of paths in $\mathbb{R}^2$.

Let us consider homotopy classes in $\SO(3)$.
Consider as in Fig. \ref{fig:2}c two loops $\Gamma_1$ and $\Gamma_2$ in $\SO(3)$ going from $\mathbb{I}_3$ to $\mathbb{I}_3$. When those loops are lifted in $\SU(2)$ as $\gamma_1$ and $\gamma_2$ starting from $\mathbb{I}_2$, all we can conclude from the construction is that the endpoints $\gamma_1(1)$ and $\gamma_2(1)$ are both $T$-preimages of $\mathbb{I}_3$.
It may happen that one of these lifted paths is a loop, i.e.~$\gamma_1(1) = \mathbb{I}_2$, while the other is an arc joining $\pm \mathbb{I}_2$, i.e.~$\gamma_2(1) = -\mathbb{I}_2$.
Both of these possibilities actually occur for loops in $\SO(3)$, and  making a continuous deformation of a loop in $\SO(3)$ does not change which behavior is seen.
Thus there are at least two types of loops in $\SO(3)$ that start at $\mathbb{I}_3$: the ones that remain loops when lifted to $\SU(2)$, and the ones that open up to arcs from $\mathbb{I}_2$ to $-\mathbb{I}_2$.

Therefore, $\SO(3)$ admits at least two homotopy classes of paths starting at $\mathbb{I}_3$, and one can show that there are in fact exactly two. On the other hand, $\SU(2)$ is the unit sphere of $\mathbb{C}^2$ (see (\ref{eq:SU2})), which is equivalently the unit sphere of $\mathbb{R}^4$.
Arguments similar to the one sketched for $\mathbb{R}^2$ above can be used to show that the unit sphere in $\mathbb{R}^{4}$ has a single homotopy class of paths (or of paths starting at a given point).

From here, it can be seen that there is no bijective continuous map from $\SU(2)$ to $\SO(3)$, as this would ultimately imply that they have the same number of homotopy classes of paths. More details and comprehensive proofs can be found in e.g.~\cite{appel_mathematics_2007}.

\section{The spinorial ball - implentation and applications}

\subsection{General principle}
\label{subsec:generalPrinc}
\begin{figure}
\centering
\includegraphics[width=15cm]{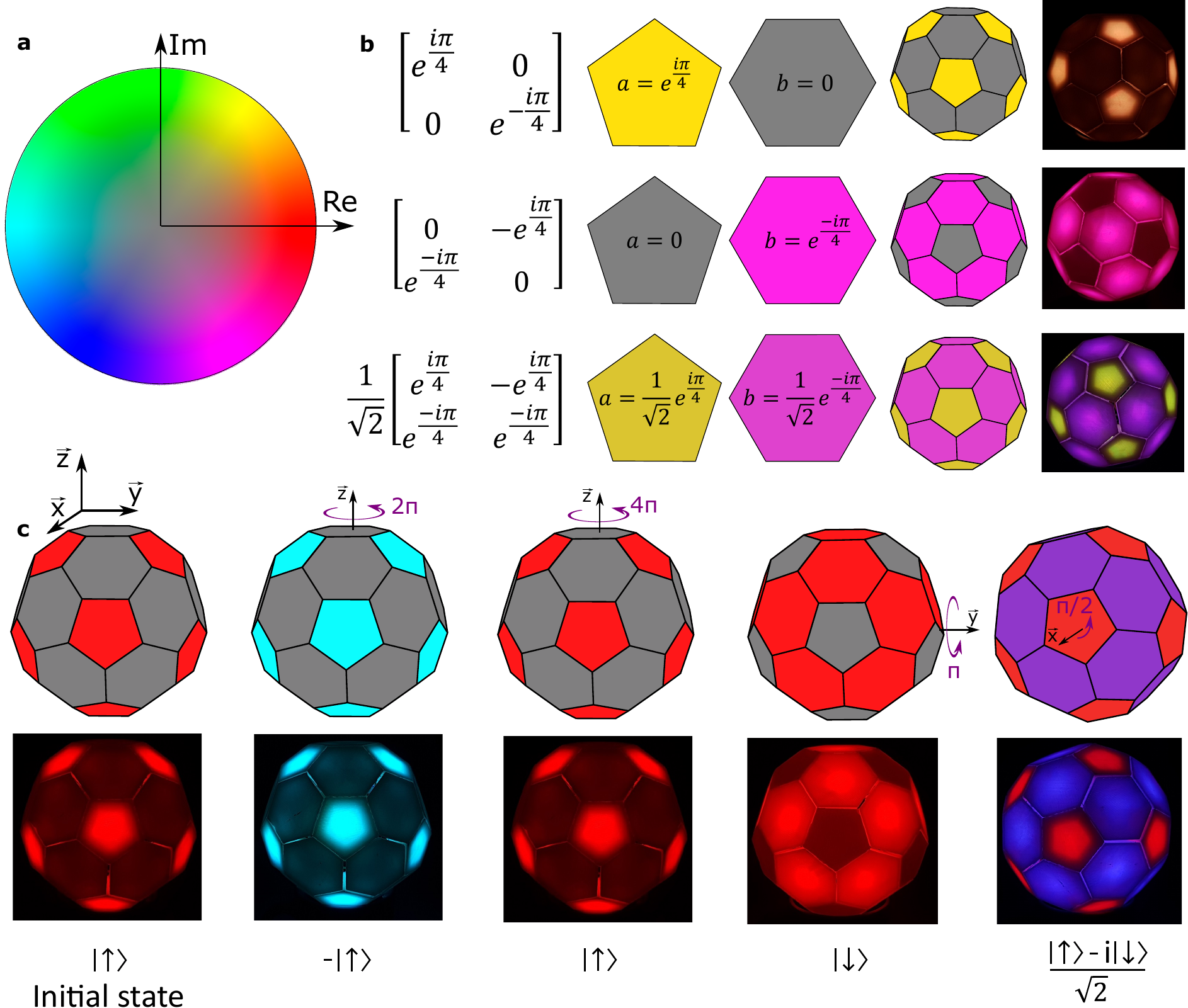}
\caption{(a) We encode a complex number using the saturation (for the modulus) and the hue (for the argument) of colors. (b) Three examples of $\SU(2)$ elements represented using two color panels and corresponding pictures of the ball. (c) Evolution of the spinorial ball starting from the state $s_0 = \ket \uparrow$ after a rotation of respectively one turn ($2 \pi)$ around $\vec z$, two turn ($4 \pi)$ around $\vec z$, half-turn ($\pi)$ around $\vec y$ and quarter-turn ($\pi/2$) around $\vec x$.}
\label{fig:3}
\end{figure}

We now have all the tools needed to fully describe what is displayed on the spinorial ball. Each complex component of a spinor $(C_P, C_H) \in \mathbb{C}^2$ is encoded as a color, with polar coordinates $(r,\theta)$ corresponding to color saturation and hue, respectively (see color map on Fig. \ref{fig:3}a).  These colors are shown on the faces, with $C_P$ as the color of the pentagons and $C_H$ the color of the hexagons. The ball is equipped with an electronic gyroscope to continuously track changes to its orientation.  It therefore knows the path of its rotations in $\SO(3)$, and lifts that path continuously to $\SU(2)$ in real time so it can display the effect of such rotations on the spinor through changes to the colors of its faces.

To make this more precise, let us assume that we initialize the ball at time $t = 0$ so its LED panels display the spinor $(1,0)$ (Fig. \ref{fig:3}c). After each time step $dt$, the gyroscope returns the current rotation matrix $R_c \in \SO(3)$ that describes how the object has been rotated since its initialization. Thus for example, if the ball has been stationary since startup, the sensor will return $\mathbb{I}_3$ after each time step. If we then start to rotate the ball and we measure the new rotation matrix after each step $dt$, we can represent each of the rotation matrices thus obtained as a small change to the previous one: From time $(k-1)\,dt$ to $k \,dt$, $R_c$ changes by an additional rotation by $\delta_k$ around $\vec{n}_k$. Thus the matrices returned by the gyroscopes are given by the time series
\begin{equation}
\begin{array}{c|ccccc}
   \textrm{Time} \quad  & 0 & dt & 2 dt & ... &  n dt \\ 
   \hline
  R_c & \mathbb{I}_3 & \quad R_{\vec{n}_1} (\delta_1)\quad &  \quad R_{\vec{n}_2} (\delta_2 )R_{\vec{n}_1} (\delta_1) \quad& ... & \quad \prod_{k=n}^1 R_{\vec{n}_k} (\delta_k) \quad \\
\end{array}
\end{equation}
As the time step $dt$ is quite small, this series is a discrete analogue of a continuous path in $\SO(3)$.
Just as we did with continuous paths, this path can be lifted to two paths in $\SU(2)$:
\begin{equation}
\begin{array}{c|ccccc}
   \textrm{Time} \quad  & 0 & dt & 2 dt & ... &  n dt \\ 
   \hline
  S_c & \mathbb{I}_2 & \quad S_{\vec{n}_1} (\delta_1)\quad &  \quad S_{\vec{n}_2} (\delta_2 )S_{\vec{n}_1} (\delta_1) \quad& ... & \quad \prod_{k=n}^1 S_{\vec{n}_k} (\delta_k) \quad \\
 & -\mathbb{I}_2 & \quad -S_{\vec{n}_1} (\delta_1)\quad &  \quad -S_{\vec{n}_2} (\delta_2 )S_{\vec{n}_1} (\delta_1) \quad& ... & \quad - \prod_{k=n}^1 S_{\vec{n}_k} (\delta_k) \quad \\
\end{array}
\label{eq:pathSU2}
\end{equation}

That is, once the ball is initialized (corresponding to choosing $\pm \mathbb{I}_2$ as a starting point), there exists a unique continuous way to transport the physical rotation of the ball to a path in $\SU(2)$. Once found, the element $S$ of $\SU(2)$ is mapped on its associated spinor $s=(C_P,C_H)$ as described in section \ref{subsec:spinSU2}) and displayed on the LED panels as illustrated in Fig. \ref{fig:3}b.

While motivated by \hspin physics, the spinorial ball can equivalently be seen  as a physical model of path lifting from $\SO(3)$ to $\SU(2)$. The LED panel shows the evolution of the spinor along the path in $\SU(2)$ when the physical ball is transformed according to a path in $\SO(3)$. The rotation the object has undergone is the image by $T$ of the element of $\SU(2)$ displayed by the LED. The symmetrical shape of the ball emphasizes this signal indication of a spinor over physical cues (such as from asymmetry) that would only show the element of $\SO(3)$.  That design element is what makes the spinorial ball behave as \hspin: Its directly evident features do not show its rotation, but rather a spinor that continuously lifts its rotation.

\subsection{Practical implementation}

The panels of the ball are 3D printed in translucent plastic and glued together in order to form a truncated icosahedron. An RGB LED is placed on the inner side of each panel. 
The inner cavity of the ball contains a battery, an orientation sensor module (Bosch BNO055), and an Arduino-compatible microcontroller board to which the sensor and all of the LEDs are connected. As is common in designs with many RGB LEDs, we use LED modules with integrated logic (WS8211 ICs) allowing them to receive color data over a serial bus, so all of the LEDs to be controlled with just three pins of the microcontroller.
It turns out that for practical reasons (computation speed or interpolation purposes, for instance), many commercial gyroscopes and orientation sensors encode rotations directly with elements of $\SU(2)$, or equivalently through quaternions, but without any guarantee that the results vary continuously.  Our discussion in section \ref{subsec:generalPrinc} is therefore slightly incorrect, as our gyroscope does not return after each time step the rotation matrix $R$ but one of the two associated elements $\pm S$ of $\SU(2)$. Depending on the previous state, the firmware of the ball will flip or not the sign of this raw spinor in order to keep the evolution of $S(t)$  continuous.  The updated spinor is then used to compute new color data for the LEDs.

All details, including the schematics, code, and 3D printing models needed to build a spinorial ball are available at \cite{gitSam}, and are free to use for any research or teaching activity.

\subsection{Classical examples}
We now give several examples of the spinorial ball evolution, the starting point always being $S_0 = \mathbb{I}_2$ corresponding to spinor $s_0 = \ket \uparrow$. The corresponding pictures of the spinorial ball after each transformation are shown in Fig. \ref{fig:3}c. 

\begin{itemize}
\item A full turn around the $\vec z$ axis (or any other axis) has the effect of moving in $\SU(2)$ from $\mathbb{I}_2$ to $-\mathbb{I}_2$, corresponding to the spinor transformation $\ket \uparrow \rightarrow - \ket \uparrow$.  The colors of the pentagonal and hexagonal faces are each inverted, i.e.~replaced with opposite hues of equal saturation.
\item Two full turns around $\vec z$ (or any other axis) restores the colors to their original state.

\item When one rotates by an angle $\phi$ around the $\vec z$ axis, one obtains spinor $\ket \uparrow \rightarrow e^{-i \phi/2} \ket \uparrow$. The phase factor $e^{-i \phi/2}$ can be observed on the spinorial ball by a color change of the pentagons while the hexagons remains dark.

\item A half-turn around the $\vec y$ axis corresponds to a path from $\mathbb{I}_2$ to $-2iS_{y}$, corresponding to $\ket \uparrow \rightarrow \ket \downarrow$. In other words, when the ball is turned upside down, the spinor goes from up-state to down-state.
\item A quarter-turn around the $\vec x$  axis corresponds to a path in $\SU(2)$ from $\mathbb{I}_2$ to $\frac{1}{\sqrt{2}} \left ( \mathbb{I}_2 -2i S_x \right )$ corresponding to $\ket \uparrow \rightarrow  \frac{1}{\sqrt{2}} \left ( \ket \uparrow - i \ket \downarrow \right )$
\end{itemize}

As mentioned earlier, the interested reader can manipulate a virtual version of the ball in our browser-based simulation using WebGL 3D graphics \citep{noauthor_httpproxy-informatiquefrquball_nodate}. A mobile version that uses a smartphone's accelerometer to sense orientation and displays the spinor as a pair of colors on the screen is also available \cite{spinPhone}.

\subsection{Visualizing homotopy classes of paths with the ball}
The homotopy classes of paths in $\SO(3)$ can also be seen with the spinorial ball. A loop in $\SO(3)$ is a rotation process that results in the same orientation of the physical ball at the end as at the beginning.
If the colors of the hexagons and pentagons displayed at the end and the beginning are the same, then the path in $\SO(3)$ lifts as a loop $\mathbb{I}_2 \rightarrow \mathbb{I}_2$ in $\SU(2)$.  But if the colors change (and are opposite according to the color map in Fig. \ref{fig:3}), the path lifts as an arc $\mathbb{I}_2 \rightarrow -\mathbb{I}_2$.  Equivalently, in terms of spins, if the spinorial ball is initialized in $\ket \uparrow$ state, then a loop of rotations corresponds to either $\ket \uparrow \rightarrow \ket \uparrow$ or $\ket \uparrow \rightarrow - \ket \uparrow$ depending on the homotopy class of the physical rotation path that has been applied. The spinorial ball can therefore be used as a homotopy class detector, similar to the Dirac belt but without any tether or constraint on its motion.

Note that the previous remark also explains why a \hspin object must come back to its initial state after two turns. After one turn, the path in $\SU(2)$ is $\mathbb{I}_2 \rightarrow -\mathbb{I}_2$. But if one performs a second turn, the corresponding path will be another arc $-\mathbb{I}_2 \rightarrow \mathbb{I}_2$. When chained, those two arcs therefore form a closed loop $\mathbb{I}_2 \rightarrow \mathbb{I}_2$ and the spin comes back to its initial state. 

\subsection{Multiply elements of $\SU(2)$ with the ball}
\label{subsec:multiplySU2}
The spinorial ball can also be used to multiply certain pairs of elements of $\SU(2)$. Let us assume that at a given time, the current rotation matrix is $R_c$ in $\SO(3)$, associated to $S_c$ in $\SU(2)$. We will now apply a small rotation $R$ associated to $\pm S$ in $\SU(2)$. The new matrix is $R \times R_c$ and as in Eq. (\ref{eq:pathSU2}), the new matrix in $\SU(2)$ is then $\pm S \times S_c$.  Here we have used that $T$ is a group homomorphism:  if it were not, lifting $R \times R_c$ might differ from lifting $R$ as $S$ and $R_c$ as $S_c$ separately and then multiplying them in $\SU(2)$.

We still need to pick a ``good'' sign for the new element of $\SU(2)$. As $R$ is close to $\mathbb{I}_3$, one of its lifts $S$ is close to $\mathbb{I}_2$ and the other ($-S$) is close from $-\mathbb{I}_2$. Since we look for continuous evolution in $\SU(2)$, the natural choice is thus $S \times S_c$, i.e.~to make $S_c$ change by as little as possible while remaining a lift of $R_c$. Thus we see that rotating the ball corresponds to multiply two elements of $\SU(2)$. For instance, rotating the ball of 180 degree around the $\vec x$ axis corresponds to multiplying the current element by $S = -2i S_x$ while a complete turn around any axis corresponds to a multiplication by $-\mathbb{I}_2$, what is consistent with previous examples. 

Finally, we mention that $\SU(2)$ can also be identified with the set of unit elements of the field of quaternions. Therefore, the ball can also be used to multiply unitary quaternions. In particular, the multiplication table of the quaternion group (a group of 8 unit quaternions) can be visualized by applying sequences of 180 degree rotations around the axis $\vec x, \vec y, \vec z$.

\subsection{Can there be a spin-1/3?}
As a last question, one may wonder whether an object might display ``spin-1/3'' behavior (or any other fractional integer spin) by using a different color encoding, or for instance by inserting a $1/3$ instead of a $1/2$ in Eq. (\ref{eq:RodrigueEuler}). If one could do so, then $\SO(3)$ would admit more than two different homotopy classes of paths (since $\mathbb{I}_2$ would admit more than two pre-images). As it is not the case, only integer or half-integer spins can exist in three dimensions. This is of course closely related to the fact that $\SU(2)$ has only one homotopy classes. The interested reader is invited to perform the experiment and explore the consequences of using $1/3$ in place of $1/2$ in formula (\ref{eq:RodrigueEuler}) to see what happens.

However, a spin-$1/3$ (or any other fractional value $p/q$) can exist in certain phenomena in two-dimensional physics \cite{PhysRevLett.48.1144}. In such cases, rotations can only occur around one axis, and the relevant set of rotations is thus the group $\SO(2)$ of matrices of the form
\begin{equation}
   R(\theta) = \begin{bmatrix}
        \cos{(\theta)} & \sin{(\theta)} \\
        -\sin{(\theta)} & \cos{(\theta)}
    \end{bmatrix}.
\end{equation}
This set of matrices is homeomorphic to the unit circle in the plane, for example by taking the first row of the matrix as the 2D coordinates of a point.  Two paths in $\SO(2)$ can then be deformed into one to another if and only if the associated paths in the unit circle have the same algebraic winding number around the origin (with each full counterclockwise turn counting as $+1$ and each clockwise turn as $-1$). The homotopy classes of paths are therefore labeled by the integers, $\mathbb{Z}$. 

Fractional spin of any integer denominator is therefore possible in this context.
Such general fractional 2D spins can be made at macroscopic scale for instance by placing a flat demonstration device on a table, so that its possible motions are rotations around the axis normal to the table, that is $\SO(2)$. The device would display a single color at any time, chosen from a circle of possible hues. When the object is physically rotated by an angle $\alpha$, the hue changes by angle $p/q \times \alpha$.  Such an object will behave as one with spin-$p/q$ in this 2D setting. In particular, it will return to its original color after $q$ full turns.

\section{Conclusion and perspectives}
In this article, we have described a macroscopic object that is able to move freely while exhibiting the all characteristics of \hspin. The latter is more than anything a geometrical feature, which turns out to have important implications in quantum mechanics. The object exhibits the connection between $\SU(2)$ and $\SO(3)$ in a visual way and shows how paths can be lifted from the latter to the former once an initial point is fixed. 

This type of object suggests several avenues for further work. First, it could be extended to any integer or half-integer spin by increasing the number of colored tiles on the ball and modifying the way they transform under rotation according to the spin matrices in the corresponding representation spaces. It would also be interesting to extend this to the case of \emph{two} macroscopic \hspin objects. Communication between their driving electronics might be used to simulate entanglement. Perhaps one could simulate the failure in Bell's inequalities in this way.

We also believe that this device could be used to popularize group theory or quantum mechanics for undergraduate students, as a practical exercise for students in electronics or engineering, or simply as an interesting toy that offers the possibility of opening a gateway to interesting mathematics and physics for any careful observer.
\section{Acknowledgment}
The authors would like to acknowledge Emmanuel Fort, Tony Jin, David Martin and Marc Abboud for insightful discussions and feedback.  Parts of this project were developed during a semester program at the Institute for Computational and Experimental Research in Mathematics (ICERM) at Brown University, and at ``Les Gustins'' Summer School with support of Jean Baud and Ingénieurs et Scientifiques de France - Sillon Alpin (IESF-SA). The authors thank these organizations and acknowledge attendees of these programs for stimulating discussions. The authors have no conflicts to disclose.
\bibliographystyle{unsrt}
\bibliography{biblio}
\end{document}